\documentclass[a4paper,12pt]{article}
\usepackage{amsmath}
\usepackage{amssymb}
\usepackage{fullpage}
\usepackage{verbatim}
	\usepackage{bbm}

\usepackage[T1,OT1]{fontenc}

\usepackage{graphicx}

\allowdisplaybreaks[1]

\usepackage{hyperref}
\usepackage{cite}

\renewcommand{\d}{\mathrm{d}}
\newcommand{\e}{\mathrm{e}}

\newcommand{\w}{\wedge}

\newcommand{\nl}{\notag \\ &\quad\,}

\begin{document}

\numberwithin{equation}{section}

\thispagestyle{empty}

\begin{flushright}
\small ITP-UH-10/14\\
\normalsize
\end{flushright}

\vspace*{0.9cm}

\begin{center}

{\LARGE \bf Curvature-induced Resolution of } 

\vskip4mm
{\LARGE \bf Anti-brane Singularities}

\vskip3mm
 {\LARGE \bf }

\vspace{1cm}
{\large Daniel Junghans${}^{1}$, Daniel Schmidt${}^{2 *}$ and Marco Zagermann${}^{2,3}$}\\

\vspace{1cm}
\vspace{.15 cm}  ${}^1$ Center for Fundamental Physics \& Institute for Advanced Study,\\
Hong Kong University of Science and Technology, Hong Kong\\

\vspace{0.5cm}
\vspace{.15 cm}   ${}^2$ {Institut f{\"u}r Theoretische Physik \&\\
${}^3$ Center for Quantum Engineering and Spacetime Research\\
Leibniz Universit{\"a}t Hannover, Appelstra{\ss}e 2, 30167 Hannover, Germany}

\vspace{1cm}
{\upshape\ttfamily daniel@ust.hk, d.schmidt@uni-jena.de, marco.zagermann@itp.uni-hannover.de}\\

\vspace{1cm}
\begin{abstract}
\noindent We study AdS$_7$ vacua of massive type IIA string theory compactified on a $3$-sphere with $H_3$ flux and anti-D$6$-branes. In such backgrounds, the anti-brane backreaction is known to generate a singularity in the $H_3$ energy density, whose interpretation has not been understood so far. We first consider supersymmetric solutions of this setup and give an analytic proof that the flux singularity is resolved there by a polarization of the anti-D$6$-branes into a D$8$-brane, which wraps a finite 2-sphere inside of the compact space. To this end, we compute the potential for a spherical probe D8-brane on top of a background with backreacting anti-D6-branes and show that it has a local maximum at zero radius and a local minimum at a finite radius of the 2-sphere. The polarization is triggered by a term in the potential due to the AdS curvature and does therefore not occur in non-compact setups where the 7d external spacetime is Minkowski. We furthermore find numerical evidence for the existence of non-supersymmetric solutions in our setup. This is supported by the observation that the general solution to the equations of motion has a continuous parameter that is suggestive of a modulus and appears to control supersymmetry breaking. Analyzing the polarization potential for the non-supersymmetric solutions, we find that the flux singularities are resolved there by brane polarization as well.
\end{abstract}
\end{center}

\setlength{\tabcolsep}{1pt}
 
\vspace{0.5cm}
\begin{center}
\footnotesize{
\begin{tabular}{l l}
$^*$&New address since March 1, 2014: Theoretisch-Physikalisches Institut, Friedrich-Schiller-Universit{\"{a}}t\\
&Jena, Fr{\"{o}}belstieg 1, 07743 Jena, Germany
\end{tabular}}
\end{center}
\setlength{\tabcolsep}{6pt}

\newpage
\tableofcontents
\vspace{0.5cm}

\section{Introduction}

Type II flux compactifications with D-brane or O-plane sources provide a very important and reasonably generic class of string backgrounds, in particular in the context of semi-realistic string model building. The best understood examples of this type are compactifications in which the flux preserves the same type of supersymmetry as the D-branes and O-planes or at least satisfies a mutual BPS-type no-force condition as in \cite{Giddings:2001yu}. The opposite case, where one combines an anti-D-brane with a flux that globally carries the charge of one or more D-branes (or vice versa), is much harder to access analytically but nevertheless of great interest, e.g., for de Sitter model building \cite{Kachru:2003aw} or holographic duals of meta-stable states in gauge theories \cite{Kachru:2002gs}.

During the past few years, various independent studies of different models of the latter type have indicated that the backreaction of oppositely charged branes (in the following simply called ``anti-branes'') induces an unusual type of singularity in their vicinity. The singularity is unusual in the sense that it occurs in the energy density of one or more $p$-form potentials that are not directly sourced by the anti-branes.
The existence of the singularity was inferred first from a computation of the linearized perturbation of the Klebanov-Strassler (KS) solution \cite{Klebanov:2000hb} by partially smeared anti-D3-branes \cite{McGuirk:2009xx, Bena:2009xk, Bena:2011hz, Bena:2011wh} (see also \cite{DeWolfe:2008zy, Dymarsky:2011pm, Massai:2012jn} for other works). An analogous singularity was then found in \cite{Blaback:2011nz, Blaback:2011pn} using non-perturbative analytical computations in a massive IIA compactification on AdS$_7\times S^3$ with $H_3$ flux and anti-D6-branes. In \cite{Bena:2012bk}, a non-perturbative proof was later also found for partially smeared anti-D3-branes in the KS background. Similar perturbative and non-perturbative results also exist for configurations with anti-M2-branes and anti-D2-branes \cite{Bena:2010gs, Giecold:2011gw, Massai:2011vi, Giecold:2013pza, Cottrell:2013asa, Blaback:2013hqa, Bena:2014bxa}, where, in contrast to their anti-D3 and anti-D6 counterparts, the singularities are not even integrable \cite{Bena:
2010gs, Giecold:2011gw, Massai:2011vi}. For the case of fully localized anti-D3-branes in the KS throat, an analytic argument for the existence of the singularity was recently given in \cite{Gautason:2013zw} based on a general relation between the classical cosmological constant and the near-brane behavior of the supergravity fields (see also \cite{Aghababaie:2003ar, Burgess:2011rv} for earlier related work and \cite{Junghans:2014xfa} for a warped effective field theory analysis of the singularity). Taken together, all these works put the existence of the flux singularities on very solid computational ground.

The physical meaning of the singular flux, on the other hand, is less clear and has been the subject of various discussions. In \cite{Blaback:2012nf}, it was suggested that the singularity might signal a perturbative instability of the backreacted solution. Another idea in the recent literature is to regulate the singularity by introducing physical IR cutoffs such as a temperature or a Hubble scale. Explicit tests showed that this is not possible in non-compact geometries \cite{Bena:2012ek, Vanriet:2013, Buchel:2013dla}, which violates a criterion for acceptable singularities due to Gubser \cite{Gubser:2000nd}.

In a different line of thought, one may wonder whether one can identify a particular stringy effect that resolves the singularity. One possibility that comes to mind is a polarization of the branes into a fuzzy higher-dimensional brane via the Myers effect \cite{Myers:1999ps}, which is known to cure singularities in the Polchinski-Strassler solution \cite{Polchinski:2000uf}. For the case of the non-compact KS background, it was indeed shown in \cite{Kachru:2002gs} that probe anti-D3-branes can polarize into an NS5-brane. Taking into account the backreaction of the anti-D3-branes, however, a polarization does not seem to occur anymore, at least not in the orthogonal D5-brane polarization channel \cite{Bena:2012vz}. Furthermore, it was shown in \cite{Bena:2012tx} that anti-D6-branes in a non-compact background with $F_0$ and $H_3$ flux do not polarize into a D8-brane. In an interesting recent paper, the analogous problem was investigated for anti-M2-branes in the CGLP background \cite{Cvetic:2000db}, the M-theory analogue of the KS solution. Using a combination of indirect arguments, the authors were able to infer the polarization potential for localized anti-M2-branes and argued that it is unstable, thus leading to a polarization of the anti-M2-branes into an M5-brane. However, a definite conclusion about the endpoint of the polarization process could not be reached, and the authors conjectured that the polarized solution is likely to be unstable itself against various decay channels.

One might therefore conclude that polarization does either not happen in solutions with anti-branes or does not lead to a (meta-)stable configuration and can therefore not resolve the singular flux. Since above results were obtained only in the context of non-compact setups, however, it is natural to ask whether compactification effects can change the conclusion. In this paper, we elaborate on this question focussing on massive type IIA flux compactifications with anti-D6-branes. For the non-compact case, their possible polarization was analyzed in \cite{Bena:2012tx} by computing the potential for a probe D8-brane that carries anti-D6-brane charge and wraps a topologically trivial $S^2$ at a distance $r$ from a large number of backreacting anti-D6-branes. The potential then turned out to have a local minimum at the position of the anti-D6-branes but no minimum away from them such that a polarization does neither happen perturbatively nor non-perturbatively via tunnelling.
The assumption of non-compactness in \cite{Bena:2012tx} was made in order to make contact with the discussion of anti-D3-branes in the non-compact KS geometry via T-duality. Moreover, it was only due to this assumption that one could actually make a definite statement about the D8-potential without knowing the entire solution of the anti-D6-brane background. Instead, owing to a universal behavior of the potential, it was sufficient to analyze the local solution in the vicinity of the anti-branes in order to rule out polarization. In the compact case, by contrast, the AdS curvature makes an extra contribution to the potential that has the right sign to turn the local minimum at the origin into a local maximum, provided the coefficients have the right magnitude. One can show that this would then also lead to the appearance of a lower-lying minimum at a finite distance away from the origin and, hence, to brane polarization. The coefficients, however, cannot be determined without further information about the global solution so that it had to be left open in \cite{Bena:2012tx} whether a polarization happens in the compact model.

The whole story received a new twist by the interesting recent work \cite{Apruzzi:2013yva}, where it was found that the anti-D6-brane model of \cite{Blaback:2011nz, Blaback:2011pn} admits solutions that preserve some supersymmetry.
This is somewhat surprising at first sight since, taking the anti-branes to be smeared along the transverse space, the very same compactifications do not seem to preserve supersymmetry in any reasonable sense \cite{Danielsson:2013qfa}. The fact that supersymmetric solutions with localized anti-branes nevertheless exist can be traced back to the possibility of having a variable Killing spinor on the 3-sphere that interpolates between the different supersymmetries preserved near and away from the anti-branes. Because of the preserved supersymmetry, the field equations are then much simpler first order equations, which facilitates the construction of numerical solutions \cite{Apruzzi:2013yva}.

In the present paper, we revisit the question of brane polarization in the compact anti-D6-brane model of \cite{Blaback:2011nz, Blaback:2011pn} in light of these new developments. In the first part of our paper, we focus on supersymmetric solutions and show that the extra constraints imposed by supersymmetry are strong enough to allow a definite statement about brane polarization. In particular, we find that the D8-brane potential has a universal form with a local maximum at the origin and a local minimum at a finite distance away from the anti-D6-branes, where the parameters of the solution can always be chosen such that the local minimum is consistent with the supergravity approximation. This means that, in the compact model, the anti-D6-branes do polarize into D8-branes and the singularity is resolved.\footnote{This seems to be consistent with independent recent findings in version 2 of \cite{Apruzzi:2013yva} regarding the D8-brane solutions themselves.} In the second part of our paper, we discuss the 
question as to whether there could also be non-supersymmetric solutions to our setup, and whether the flux singularity is resolved there by brane polarization as well. In this paper, we give numerical evidence for a one-parameter family of solutions to the full second order field equations, which are only supersymmetric for a special choice of the parameter. Computing the D8-brane potential, we find that different regimes of this parameter lead to a different qualitative behavior. For non-supersymmetric solutions that are close to the supersymmetric one in moduli space, we find a similar behavior with a local maximum at the origin and a local minimum at a finite distance away from it. The mass of the worldvolume scalar responsible for brane polarization then satisfies the Breitenlohner-Freedman (BF) bound \cite{Breitenlohner:1982bm, Breitenlohner:1982jf} such that polarization happens non-perturbatively in this regime. Further away from the supersymmetric point, however, we also find regimes that are physically different. Some of the non-supersymmetric solutions are then tachyonic, i.e., the mass of the worldvolume scalar sinks below the BF bound such that brane polarization can already happen perturbatively.

This paper is organized as follows. In Section \ref{sec:setup}, we present the setup of our model, review some earlier results used in this paper and discuss the parameter space of the general solution. In Section \ref{sec:potential}, we discuss the polarization potential and its regime of validity. In Section \ref{sec:susy}, we analyze supersymmetric solutions of our setup and give an analytic proof that the flux singularity is resolved there by brane polarization. In Section \ref{sec:non-susy}, we present numerical evidence for the existence of non-supersymmetric solutions and show that the anti-D6-branes polarize there as well. We conclude in Section \ref{sec:conclusions} with a summary of our results and some interesting questions for future research. For some of the technical details and further background material on the numerical treatment used in this work, we refer to \cite{Schmidt:2013}.

\section{General setup}
\label{sec:setup}

\subsection{Ansatz and field equations}

We consider compactifications of massive type IIA supergravity on AdS$_7 \times S^3$ with spacetime filling anti-D6-branes and $H_3$ flux 
threading the $3$-sphere. This model was first proposed in \cite{Blaback:2010sj} and analyzed in detail in \cite{Blaback:2011nz, Blaback:2011pn}. 
If one smears the anti-D$6$-branes across the $S^3$, 
it is straightforward to find global solutions to the equations of 
motion \cite{Blaback:2010sj, Blaback:2011nz}. These solutions are non-supersymmetric \cite{Danielsson:2013qfa} but nevertheless perturbatively 
stable in the sector of the left-invariant deformations \cite{Blaback:2011nz}. 
The broken supersymmetry is plausible because the $H_3$ flux has to carry D$6$-brane 
charge in order to cancel the global tadpole of the anti-D6-branes.

Treating the anti-D6-branes as localized objects that are pointlike in the compact dimensions, the solution aquires a warp and a conformal factor, as well as non-trivial dilaton and $F_2$ profiles \cite{Blaback:2011nz}. Furthermore, contrary to the smeared case, supersymmetric solutions are possible \cite{Apruzzi:2013yva}. In order to simplify the equations of motion, we assume that all anti-D6-branes are sitting on either of the two poles of the 3-sphere such that an SO$(3)$ rotational symmetry is preserved and the system becomes effectively one-dimensional \cite{Blaback:2011nz}. Using standard spherical coordinates, we denote by $\theta\in [0,\pi] $ the angular coordinate that interpolates between the north pole and the south pole of the 3-sphere and parameterize the metric as
\begin{equation}
\d s^2_{10}=\e^{2A(\theta)} \d s^2_{\textrm{AdS}_7}+\e^{2B(\theta)}\left( \d \theta^2+\sin^2(\theta) \d s_{S^2}^2 \right), \label{e:metric_ansatz}
\end{equation}
where 
\begin{equation}
\d s^2_{S^2}=\d \varphi^2+\sin^2(\varphi) \d \chi^2 \label{e:s2_lineele}
\end{equation}
is the line element of an ordinary 2-sphere and $\d s^2_{\textrm{AdS}_7}$ the one for a 7-dimensional Anti-de Sitter space. Here, we have included a warp factor $A$ and a conformal factor $B$, which can depend on the coordinate $\theta$.\footnote{The SO$(3)$ invariance would also allow a separate conformal factor in front of the $S^2$ line element. With a redefinition of $\theta$, however, this can always be absorbed such that the form \eqref{e:metric_ansatz} is obtained.}

The most general ansatz for the fluxes compatible with the symmetries of our setup is
\begin{align}
H &= \lambda(\theta) F_0 \e^{\frac{7}{4}\phi(\theta)} \star_3 \mathbbm{1}, \label{e:H_ansatz}\\
F_2 &= \e^{-\frac{3}{2}\phi(\theta)-7A(\theta)}\star_3 \d \alpha(\theta), \label{e:F2_ansatz}
\end{align}
where $\phi(\theta)$ denotes the dilaton and we  introduced  two functions $\alpha( \theta)$ and $\lambda( \theta )$ as in  \cite{Blaback:2011nz}. Using the $H_3$ equation of motion, one finds an algebraic relation between the functions $\lambda(\theta)$ and $\alpha(\theta)$,
\begin{equation}
\alpha + \textrm{const}=\e^{\frac{3}{4}\phi+7A}\lambda. \label{e:H_eom}
\end{equation}
The additive constant can always be absorbed into a redefined $\alpha$, which we will assume in the following. We thus end up with four independent functions of one variable: $A(\theta)$, $B(\theta)$, $\phi(\theta)$ as well as either $\lambda(\theta)$ or $\alpha(\theta)$.

The non-trivial equations of motion are the $F_2$ Bianchi identity, the dilaton equation, the trace of the external Einstein equation as well as the internal Einstein equations along the $(\theta\theta)$-direction and along the transverse directions. They read (in this order) \cite{Blaback:2011nz}:
\begin{align}
0 &= - \frac{\left({\e^{-\tfrac{3}{2}\phi-7A+B} \sin^2 \theta\, \alpha^\prime}\right)^\prime}{\e^{3B} \sin^2 \theta} + \e^{\tfrac{7}{4}\phi} \lambda F_0^2 + Q \delta(\Sigma), \label{eoms-theta-bianchi} \\
0 &= - \frac{\left({\e^{7A+B} \sin^2 \theta\, \phi^\prime}\right)^\prime}{\e^{7A+3B} \sin^2 \theta} + \e^{\tfrac{5}{2}\phi} F_0^2 \left({\frac{5}{4} - \frac{\lambda^2}{2}}\right) + \frac{3}{4} \e^{-14A-2B-\tfrac{3}{2}\phi} \left({\alpha^\prime}\right)^2 + \frac{3}{4} \e^{\tfrac{3}{4}\phi} T \delta(\Sigma), \label{eoms-theta-dilaton} \\
0 &= 96 \e^{-2A} + 16 \e^{-2B} \left[{7 \left({A^\prime}\right)^2 + A^\prime B^\prime + \frac{\left({\sin^2 \theta\, A^\prime}\right)^\prime}{\sin^2 \theta}}\right] + \e^{\tfrac{5}{2}\phi} F_0^2 \left({1 - 2\lambda^2}\right) \nl - \e^{-14A-2B-\tfrac{3}{2}\phi} \left({\alpha^\prime}\right)^2 - \e^{\tfrac{3}{4}\phi} T \delta(\Sigma), \label{eoms-theta-exteinstein} \\
0 &= - 2 + \frac{\left({\sin^2 \theta \, B^\prime}\right)^\prime}{\sin^2 \theta} + 7 \left({A^\prime}\right)^2 + B^{\prime\prime} + 7 A^{\prime\prime} - 7 A^\prime B^\prime \nl +\frac{1}{2} \left({\phi^\prime}\right)^2 + \frac{1}{16} \e^{\tfrac{5}{2}\phi+2B} F_0^2 \left({1+6\lambda^2}\right) - \frac{1}{16} \e^{-14A-\tfrac{3}{2}\phi} \left({\alpha^\prime}\right)^2 + \frac{7}{16} \e^{\tfrac{3}{4}\phi+2B}T\delta(\Sigma), \label{eoms-theta-inteinsteintheta} \\
0 &= - 2 + \frac{\left({\sin^2 \theta \, B^\prime}\right)^\prime}{\sin^2 \theta} + \left({B^\prime}\right)^2 + \cot \theta \left({7A+B}\right)^\prime + 7 A^\prime B^\prime \nl + \frac{1}{16} \e^{\tfrac{5}{2}\phi+2B} F_0^2 \left({1+6\lambda^2}\right) + \frac{7}{16} \e^{-14A-\tfrac{3}{2}\phi} \left({\alpha^\prime}\right)^2 + \frac{7}{16} \e^{\tfrac{3}{4}\phi+2B}T\delta(\Sigma), \label{eoms-theta-inteinsteintrans}
\end{align}
where primes are derivatives with respect to $\theta$ and $\delta(\Sigma)$ should be read as a sum of delta distributions due to the localized sources at the north and south pole. One can verify that all other equations of motion are automatically satisfied for the above ansatz of the fields. Note that, while \eqref{eoms-theta-bianchi}--\eqref{eoms-theta-inteinsteintrans} seem to imply five second order ODEs for the four functions $A$, $B$, $\phi$ and $\alpha$, only four of them are really independent.\footnote{The three Einstein equations can be combined to give a constraint equation that only contains first derivatives of $A$, $B$, $\phi$ and $\alpha$. Taking the derivative of this constraint can then be used to derive, e.g., the second order equation for $\alpha$.}

\subsection{Near-brane expansion}

Although the above ansatz considerably simplifies the equations of motion, finding the general analytic solution to the differential equations \eqref{eoms-theta-bianchi}--\eqref{eoms-theta-inteinsteintrans} is still a difficult problem. In \cite{Blaback:2011pn}, however, it was noted that, even in the absence of the full solution, one can obtain useful information by performing an expansion of the fields $A$, $B$, $\phi$ and $\lambda$ around the north pole $\theta=0$ (or, equivalently, the south pole $\theta=\pi$). Solving the equations of motion order by order in this expansion, one finds surprisingly strong constraints on the behavior of the fields. In particular, it was shown in \cite{Blaback:2011pn} that only two different boundary conditions at the pole are consistent with the equations of motion. The first boundary condition describes fields in the vicinity of the anti-D6-branes\footnote{Instead of the anti-D6-branes considered in this paper, also D6-branes are allowed as consistent sources.}, while the second boundary condition is valid for a pole without any localized sources. The small-$\theta$ behavior of the fields for a pole with anti-branes is then given by
\begin{gather}
\e^{-A(\theta)} = \theta^{-\frac{1}{16}} \left( a_0 + a_1 \theta + a_2 \theta^2 + \ldots \right), \quad \e^{-2B(\theta)} = \theta^{\frac{7}{8}} \left( b_0 + b_1 \theta + b_2 \theta^2 + \ldots \right), \qquad\,\, \notag \\ \e^{-\tfrac{1}{4}\phi(\theta)} = \theta^{-\frac{3}{16}} \left( f_0 + f_1 \theta + f_2 \theta^2 + \ldots \right), \quad \lambda(\theta) = \theta^{-1} \left( \lambda_0 + \lambda_1 \theta + \lambda_2 \theta^2 + \ldots \right), \label{eq:d6-ansatz-bc}
\end{gather}
where $a_i$, $b_i$, $f_i$ and $\lambda_i$ are certain expansion coefficients that we will discuss momentarily. For a pole without any sources, one finds a non-singular behavior of the fields,
\begin{gather}
\e^{-A(\theta)} = \tilde a_0 + \tilde a_2 \theta^2 + \tilde a_4 \theta^4 + \ldots, \quad \e^{-2B(\theta)} = \tilde b_0 + \tilde b_2 \theta^2 + \tilde b_4 \theta^4 + \ldots, \notag \\ \e^{-\tfrac{1}{4}\phi(\theta)} = \tilde f_0 + \tilde f_2 \theta^2 + \tilde f_4 \theta^4 + \ldots , \quad \lambda(\theta) = \tilde \lambda_0 + \tilde \lambda_2 \theta^2 + \tilde \lambda_4 \theta^4 + \ldots, \label{eq:d6-ansatz-bc2}
\end{gather}
where $\tilde a_i$, $\tilde b_i$, $\tilde f_i$ and $\tilde \lambda_i$ are again some expansion coefficients, which are non-zero only for even powers of $\theta$ due to symmetry reasons. Note that an analogous expansion of the fields around the south pole $\theta=\pi$ can always be obtained by replacing $\theta \to \pi-\theta$ in above equations. By using the metric ansatz \eqref{e:metric_ansatz} together with the near-brane boundary condition \eqref{eq:d6-ansatz-bc} in \eqref{e:H_ansatz}, one finds that the energy density of the $H_3$ flux is divergent at the anti-brane position \cite{Blaback:2011pn},
\begin{equation}
\e^{-\phi} |H|^2 \propto \theta^{-\frac{1}{8}}. \label{singularity}
\end{equation}
The resolution of this singularity by brane polarization is the topic of this paper.

\subsection{Parameter space of the general solution}
\label{sec:parameters}

Substituting the small-$\theta$ solution \eqref{eq:d6-ansatz-bc} into the equations of motion, one finds that it has 6 free parameters, which are given by the lowest order expansion coefficients $a_0$, $b_0$, $f_0$, $\lambda_0$, $\lambda_1$ and the Romans mass $F_0$. All higher order coefficients $a_i$, $b_i$, $f_i$, $\lambda_i$ are fixed in terms of these parameters, as can be checked by solving the equations of motion order by order in the $\theta$ expansion \cite{Blaback:2011pn}. In order to obtain some information about the global solution from our knowledge of the local solution, it is crucial to understand the physical meaning of the 6-dimensional parameter space. Let us therefore explain the origin of these parameters in detail.

We first discuss the parameters $b_0$, $f_0$ and $F_0$, whose interpretation is the easiest:

\begin{itemize}
\item One combination of the parameters fixes the charge $Q_1$ of the source that sits at the north pole. This combination can explicitly be determined by an analysis of the divergent field behavior near the localized sources. For the case of anti-D6-branes, the charge is negative and given by $Q_1 = - f_0^3/\sqrt{b_0}$ \cite{Blaback:2011pn}.\footnote{In the special case $Q_1=0$ where no source sits at the pole, the correct behavior of the fields is not given by \eqref{eq:d6-ansatz-bc} but by the smooth boundary condition \eqref{eq:d6-ansatz-bc2}. Choosing the latter on the north pole then automatically restricts to the $Q_1=0$ subspace of the general solution without fixing any of the expansion coefficients $\tilde a_i, \tilde b_i, \tilde f_i, \tilde \lambda_i, F_0$. This is consistent with the fact that substituting \eqref{eq:d6-ansatz-bc2} instead of \eqref{eq:d6-ansatz-bc} into the equations of motion yields only 5 instead of 6 free parameters.}

\item The parameter $F_0$ fixes the Romans mass.

\item One combination of the parameters is related to a residual gauge degree of freedom that is not fixed in our ansatz for the internal metric \eqref{e:metric_ansatz}. This ansatz still allows for a redefinition $\theta \to \tilde\theta(\theta)$ of the spherical coordinates, as can be seen by considering coordinate transformations of \eqref{e:metric_ansatz} satisfying
\begin{equation}
\e^{2 \tilde B(\tilde \theta)} \d \tilde \theta^2 = \e^{2 B(\theta)} \d \theta^2, \qquad \e^{2 \tilde B(\tilde \theta)} \sin ^2 (\tilde \theta) = \e^{2 B(\theta)} \sin ^2 (\theta).
\end{equation}
Combining the two conditions yields the ODE
\begin{equation}
\frac{\d \tilde \theta(\theta)}{\d \theta} = \frac{\sin \tilde \theta}{\sin \theta},
\end{equation}
which can be solved to find a one-parameter family of solutions for $\tilde \theta(\theta)$ of the form $\tilde \theta(\theta) = \textrm{const} \cdot \theta + \mathcal{O}(\theta^2)$ (cf. a similar discussion in \cite{Blaback:2011nz}). This reparametrization freedom can be fixed by setting the parameter $b_0$ to an arbitrary value.
\end{itemize}

The remaining 3 parameters $a_0$, $\lambda_0$ and $\lambda_1$ depend on global properties of the solution such that we were not able to analytically determine how their deformation affects the solution. However, as mentioned in the introduction, we also looked for numerical solutions of the equations of motion \eqref{eoms-theta-bianchi}--\eqref{eoms-theta-inteinsteintrans}. An explicit check of the properties of the numerical solutions in different regions of the parameter space then lead to the following interpretation of $a_0$, $\lambda_0$ and $\lambda_1$:

\begin{itemize}
\item One combination of the parameters fixes the charge $Q_2$ of a possible source located at the south pole, i.e., the pole opposite to the pole around which we expand the fields. In order to explicitly determine the combination of parameters that equals $Q_2$, one would have to connect the small-$\theta$ expansion of the fields to the field behavior near the opposite pole, which is not possible in the absence of a full analytic solution. We have therefore not been able to find an analytic expression for $Q_2$ in terms of the local parameters $\left\{a_0, b_0, f_0, \lambda_0, \lambda_1, F_0\right\}$. However, our numerical simulations verify that indeed one direction in the full 6-dimensional parameter space controls the value of the charge at the south pole.

\item One combination of the parameters is related to the integration range of the fields $A(\theta)$, $B(\theta)$, $\phi(\theta)$ and $\lambda(\theta)$, i.e., the range $\theta \in [0, \theta_\textrm{int}]$ within which none of the fields diverge. Since the compact space in our setup is a conformal 3-sphere and our ansatz allows the presence of sources only at $\theta=0$ and/or $\theta=\pi$, we demand that our solutions have an integration range $\theta_\textrm{int} = \pi$, i.e., the fields are only allowed to diverge at either of the two poles but have to be regular inbetween.\footnote{We also found numerical solutions with a smaller integration range some of which may have a physical interpretation as well. We leave the discussion of these solutions for future work.} Note that, contrary to an assumption in \cite{Danielsson:2013qfa}, our numerical simulations show that this requirement fixes only one direction in the 6-dimensional parameter space.

\item One combination of the parameters, which we denote by $\sigma$ in the following, appears to be a modulus. As we will discuss in Section \ref{sec:non-susy}, it is related to supersymmetry breaking. This interpretation is again supported by our numerical simulations, which show that the supersymmetry conditions are only fulfilled for certain values of this parameter, while the general equations of motion are satisfied to good accuracy for all other values as well. Just like for the two parameters controlling $Q_2$ and $\theta_\textrm{int}$, we were not able to find an analytic expression of $\sigma$ in terms of the local parameters $\left\{a_0, b_0, f_0, \lambda_0, \lambda_1, F_0\right\}$. We will show in Section \ref{sec:susy}, however, that the requirements of supersymmetry and integration range $\theta_\textrm{int}=\pi$ taken \emph{together} can explicitly be written as two conditions for $\lambda_0$ and $\lambda_1$ in terms of the other local parameters.

\end{itemize}

To summarize, we have used a combination of analytic and numerical insights to conclude that, after fixing a gauge for $\theta$ and restricting to solutions that integrate to $\pi$, the general solution to \eqref{eoms-theta-bianchi}--\eqref{eoms-theta-inteinsteintrans} has one continuous parameter $\sigma$, which is suggestive of a modulus, and 3 discrete parameters $\left\{Q_1,Q_2,F_0\right\}$, which specify the brane charges at the two poles as well as the flux number.

\section{Brane polarization}
\label{sec:potential}

\subsection{The D8-brane potential}

As discussed in the introduction, the Myers effect \cite{Myers:1999ps} provides a possible mechanism for the resolution of the unusual anti-brane singularity \eqref{singularity}. According to this idea, the backreacted field configuration near the anti-D6-branes could trigger their polarization into a D8-brane, which wraps a topologically trivial $S^2$ at a finite $\theta=\theta_\star$ away from the original position of the branes and has the anti-D6 charge dissolved into worldvolume flux. The original solution would then have to be cut off at $\theta=\theta_\star$ and matched to a non-singular solution in the interior of the D8-brane, so that the singularity disappears. For this to be possible, the configuration with the D8-brane at $\theta_\star>0$ has to be dynamically favored. One way to test this is to consider a probe D8-brane carrying a large number $n$ of anti-D6-brane charge and place it into the backreacted field configuration sourced by an even larger number $N \gg n$ of anti-D6-branes \cite{Bena:
2012tx}. The DBI and WZ action of the D8-brane then induce an effective potential $V(\theta)$ for the D8-brane position, which, in order that brane polarization occurs, should have a local extremum at $\theta=0$ and a lower-lying local mimimum at some finite $\theta=\theta_\star>0$. The purpose of the present section is to compute this potential, where we will closely follow the analysis of \cite{Bena:2012tx}.

We start with the D$8$-brane action in Einstein frame,
\begin{equation}
S_{\textrm{D}8} = - \mu_8 \int \d^9 \xi\, \e^{\tfrac{5}{4}\phi} \sqrt{-\det \big( g_{\alpha\beta} - \e^{-\phi/2} \mathcal{F}_{\alpha\beta}\big)} + \mu_8 \int \left({C_9 - \mathcal{F} \w C_7}\right), \label{nc-d8brane-action1}
\end{equation}
where $\mathcal{F} = B + 2\pi F$ in string units and $F$ is the worldvolume gauge field strength. The latter can be determined by demanding that the WZ term in the action carries $n$ units of anti-D6-brane charge. This yields
\begin{equation}
F = \frac{n}{2} \mathrm{vol}_{S^2}, \label{nc-fluxansatz2}
\end{equation}
where $\mathrm{vol}_{S^2}=\sin(\varphi)\d\varphi \w \d\chi$ is the volume-form of the 2-sphere.
The gauge potentials appearing in the action are given by $\d B = H$, $\d C_7 = - \e^{\frac{3}{2} \phi} \star_{10} F_2$ and $\d C_9 = \e^{\frac{5}{2}\phi} \star_{10} F_0 + H \wedge C_7$ in our conventions and can be expressed in terms of the fields $A$, $B$, $\phi$ and $\lambda$ by using these definitions together with our ansatz \eqref{e:metric_ansatz}, \eqref{e:H_ansatz} and \eqref{e:F2_ansatz}.

We can then substitute these expressions into \eqref{nc-d8brane-action1} and perform a double expansion of the action in powers of $1/n$ and $\theta$ using \eqref{eq:d6-ansatz-bc} (we refer to \cite{Bena:2012tx} for more details on the computation). The regime in which this expansion is consistent will be discussed below. One finds that it is sufficient to
consider the three leading terms in the potential \cite{Bena:2012tx},
\begin{equation}
V(\theta) \propto \frac{n}{2} c_2 \theta^2 - c_3 \theta^3 + \frac{2}{n} c_4 \theta^4, \label{d8potential}
\end{equation}
where the coefficients are given by\footnote{By flipping the orientation of the D8-brane, it is always possible to change the sign of the coefficient in front of the cubic term. In the following, we will restrict to one choice for this sign, as this is sufficient to show that polarization occurs in our solutions.}
\begin{equation}
c_2 = - \frac{7}{a_0^5 b_0 f_0^3} + \frac{1}{12} \frac{\lambda_0^2 F_0^2}{a_0^7 b_0 f_0^{13}}, \quad c_3 = \frac{1}{3} \frac{ \lambda_0 F_0}{a_0^7 b_0^{3/2} f_0^{10}}, \quad c_4 = \frac{1}{2} \frac{1}{a_0^7 b_0^2 f_0^{7}}. \label{pol-coefficients}
\end{equation}
The important point to note here is that the potential \eqref{d8potential} favors brane polarization if the coefficient $c_2$ in front of the quadratic term is negative since the potential then has a local minimum at finite $\theta$ that is lower than the local maximum at the origin. If, on the other hand, $c_2$ is positive, it depends on its magnitude whether or not such a minimum at finite $\theta$ exists. Without further knowledge of the parameters $a_0$, $b_0$, $f_0$, $\lambda_0$ and $F_0$, it would thus be impossible to decide whether the branes polarize or not. This is different in the non-compact model studied in \cite{Bena:2012tx}, where the external spacetime was chosen to be Minkowski. The first term in $c_2$, which can be traced back to the AdS curvature, is then absent such that $c_2$ is always positive. Furthermore, one can then check that there is no other minimum away from $\theta=0$ such that polarization can neither happen perturbatively nor non-perturbatively. For the non-compact model, one therefore concludes that the anti-D6-branes do not polarize and the singularity prevails \cite{Bena:2012tx}. In Sections \ref{sec:susy} and \ref{sec:non-susy}, we will feed in additional information on the parameters $a_0$, $b_0$, $f_0$, $\lambda_0$ and $F_0$ that will allow us to determine the sign and magnitude of $c_2$ also for the compact model.

An important question is furthermore whether, for $c_2 < 0$, the mass of the worldvolume scalar $\theta$ is above or below the BF bound \cite{Breitenlohner:1982bm, Breitenlohner:1982jf}. Computing the kinetic term for $\theta$ from the DBI action, we find that the Lagrangian to quadratic order takes the form
\begin{equation}
\mathcal{L}(\theta) \propto - \frac{1}{2 a_0^5 b_0 f_0^{3}} ({\partial_\mu \theta})^2 - c_2 \theta^2 + \mathcal{O}(\theta^3),
\end{equation}
where the kinetic term is contracted with the unwarped AdS$_7$ metric, which has unit radius in our conventions. Note that the overall proportionality factor of the Lagrangian does not matter at this order as it can always be absorbed by a field redefinition. We thus find that the mass of the scalar is given by
\begin{equation}
m^2 = 2c_2 a_0^5 b_0 f_0^{3} = -14 + \frac{1}{6} \frac{\lambda_0^2 F_0^2}{a_0^2 f_0^{10}}. \label{mass}
\end{equation}
For a canonically normalized scalar field in $(d+1)$-dimensional AdS space with unit radius, the BF bound is $m^2 \ge - \frac{d^2}{4}$, which in our case becomes $m^2 \ge - 9$.

\subsection{Regime of validity}
\label{regime}

In order to derive the potential \eqref{d8potential}, we used several approximations. A simple way to show that these are justified is to consider scaling symmetries of the supergravity equations \cite{Witten:1985xb, Burgess:1985zz} (see also \cite{Burgess:2011rv, Gautason:2013zw}). In particular, one verifies that \eqref{eoms-theta-bianchi}--\eqref{eoms-theta-inteinsteintrans} are invariant under the global rescalings 
\begin{gather}
\e^{-A} \to \zeta^{3/8} \xi^{-5/8} \e^{-A}, \quad \e^{-2B} \to \zeta^{3/4} \xi^{-5/4} \e^{-2B}, \quad \e^{-\frac{1}{4}\phi} \to \zeta^{1/8} \xi^{1/8} \e^{-\frac{1}{4}\phi}, \notag \\ \lambda \to \lambda, \quad F_0 \to \zeta F_0, \quad Q \to \xi Q, \quad T \to \xi T.
\end{gather}
The expansion \eqref{eq:d6-ansatz-bc} and the fact that $|Q| = T = N \mu_6$ for a background with $N$ anti-D6-branes then imply that the expansion coefficients scale like
\begin{equation}
a_0 \sim F_0^{3/8} N^{-5/8}, \quad b_0 \sim F_0^{3/4} N^{-5/4}, \quad f_0 \sim F_0^{1/8} N^{1/8}, \quad \lambda_0 \sim F_0^{0} N^{0}. \label{scale}
\end{equation}
Using these scalings in \eqref{pol-coefficients}, we furthermore find
\begin{equation}
c_2 \sim F_0^{-3} N^4, \quad c_3 \sim F_0^{-4} N^5, \quad c_4 \sim F_0^{-5} N^6.
\end{equation}
We can thus hope that, by choosing an appropriate regime for $n$, $N$ and $F_0$, we can rescale the terms in the potential \eqref{d8potential} such that our different approximations are all satisfied at the same time. That this is indeed the case can be seen by checking the following conditions \cite{Bena:2012tx}:
\begin{itemize}
\item We consider $n$ probe D$6$-branes that polarize into a D$8$-brane via the Myers effect. This description is valid if
\begin{equation}
1 \ll n \ll N.
\end{equation}
\item The radius of the D$8$-brane at the minimum of the potential, $\theta_\star = n \frac{3c_3 \pm \sqrt{9c_3^2 - 32c_2c_4}}{16 c_4}$, must be small in order that the $\theta$ expansion is still valid. Hence,
\begin{equation}
n \frac{F_0}{N} \ll 1. 
\end{equation}
\item The expansion of the DBI action in powers of $1/n$ is justified if $\det (\e^{\phi/2} g^{\scriptscriptstyle{(S^{\scriptscriptstyle{2}})}}_{\alpha\beta}) \ll \det \mathcal{F}_{\alpha\beta}$, where $g^{\scriptscriptstyle{(S^{\scriptscriptstyle{2}})}}_{\alpha\beta}$ denotes the metric along the 2-sphere wrapped by the D$8$-brane. This implies $n \gg \frac{\theta_\star^{3/2}}{f_0^2b_0}$, which again yields
\begin{equation}
n \frac{F_0}{N} \ll 1. 
\end{equation}
\item The radius of the 2-sphere wrapped by the D$8$-brane should be large in string units, $\det (\e^{\phi/2} g^{\scriptscriptstyle{(S^{\scriptscriptstyle{2}})}}_{\alpha\beta}) \gg 1$. This yields $1 \ll \frac{\theta_\star^{3/2}}{f_0^2b_0}$ and, hence,
\begin{equation}
n  \gg \left(\frac{N}{F_0}\right)^{1/3}.
\end{equation}
This condition ensures that the background curvature is small at $\theta_\star$.
\item The string coupling $\e^\phi$ should be small at the minimum, i.e., $f_0 \theta_\star^{-3/16} \gg 1$. This leads to the condition
\begin{equation}
n  \ll \left(\frac{N^5}{F_0}\right)^{1/3}.
\end{equation}
\end{itemize}
These conditions agree exactly with the conditions for the non-compact model, which were obtained in \cite{Bena:2012tx} using a somewhat different reasoning. It is straightforward to check that all conditions can be satisfied together, e.g., for the choice $n=20$, $N = 400$ and $F_0 = 4$.

\section{The supersymmetric solution}
\label{sec:susy}

In the recent work \cite{Apruzzi:2013yva}, supersymmetric warped compactifications of the form AdS$_7\times \mathcal{M}_3$ of the type II supergravity theories were classified. For the case of massive type IIA supergravity considered in the present paper, it was found in \cite{Apruzzi:2013yva} that a compact $\mathcal{M}_3$ must have $S^3$ topology, and numerical solutions to the corresponding supersymmetry equations were presented. These solutions must therefore be contained in the framework studied in Section \ref{sec:setup} and in the earlier works \cite{Blaback:2011nz, Blaback:2011pn}. In this section, we translate the supersymmetric solutions of \cite{Apruzzi:2013yva} into our language and identify the constraint hypersurface they correspond to in our parameter space $\left\{a_0, b_0, f_0, \lambda_0, \lambda_1, F_0\right\}$ (see also \cite{Schmidt:2013}). This will allow us to make a definite statement about the sign of the coefficient $c_2$ in the potential \eqref{d8potential} for this class of compact models.

\subsection{The supersymmetry conditions}

In the notation used in \cite{Apruzzi:2013yva}, the metric is given in string frame and reads
\begin{equation}
\d s_{10\, \textrm{string}}^2=\e^{2\tilde A(r)}\, \d s^2_{\textrm{AdS}_7}+ \d s^2_{\mathcal{M}_3},
\end{equation}
where
\begin{equation}
\d s^2_{\mathcal{M}_3}=\d r^2+\frac{1}{16}\e^{2\tilde{A}(r)}\left( 1-x(r)^2 \right)\d s^2_{S^2}\label{e:susy_metric}
\end{equation}
takes the form of an $S^2$ fibration over a compact interval $r\in[r_n,r_s]$, whose boundaries are defined by the vanishing of the $S^2$ radius, i.e., the points $r$ at which $x(r)^2=1$, corresponding to the two poles of the 3-sphere. The $H_3$ and $F_2$ field strengths are given by
\begin{align}
H&= -\left( 6\e^{-\tilde A(r)}+x(r)  F_0 \e^{\tilde\phi(r)} \right)\,\mathrm{vol}_{S^3}, \label{Hdef}\\
F_2&=\frac{1}{16}\sqrt{1-x(r)^2}\e^{\tilde A(r)-\tilde \phi(r)}\left( x(r) \e^{\tilde A(r)+\tilde\phi(r)} F_0-4 \right)\mathrm{vol}_{S^2}. \label{Fdef}
\end{align}
Converting the metric to Einstein frame and comparing with our ansatz \eqref{e:metric_ansatz}, one can express the functions $\tilde{A}(r)$, $\tilde{\phi}(r)$ and $x(r)$ in terms of our functions $A(\theta)$, $B(\theta)$ and $\phi(\theta)$:
\begin{align}
\tilde A(r(\theta))&=A(\theta)+\frac{1}{4}\phi(\theta), \label{e:A_relation}\\
\tilde{\phi}(r(\theta))&=\phi(\theta), \label{e:phi_relation}\\
x(r(\theta))^2&=1-16\, \e^{2 B(\theta)-2 A(\theta)}\sin^2(\theta). \label{e:x_relation}
\end{align}
Moreover, we obtain the relation
\begin{equation}
\left(\frac{\d \theta}{\d r}\right)^2=\frac{16\sin^2\left( \theta\left( r \right) \right)}{\left( 1-x( r)^2 \right)}\e^{-2\tilde A(r)} = \e^{-2 B\left( \theta \right)-\frac{\phi\left( \theta \right)}{2}} \label{e:rt_dgl_r}
\end{equation}
and, by comparing (\ref{Hdef}) and (\ref{Fdef}) with our flux ansatz (\ref{e:H_ansatz}) and (\ref{e:F2_ansatz}),
\begin{align}
\lambda(\theta) & = -\frac{1}{F_0}\e^{-\phi(\theta)}\left( 6\e^{-\tilde A(r(\theta))}+x(r(\theta)) F_0\e^{\tilde\phi(r(\theta))} \right), \label{e:alpha_r_H} \\
\partial_\theta \alpha (\theta) & = \frac{1}{4 \sin(\theta)} \e^{\frac{7}{4}\phi(\theta)+7A(\theta)} \left( x(r(\theta)) \e^{\tilde A(r(\theta))} F_0  - 4 \e^{- \tilde\phi(r(\theta))}\right). \label{e:alphaprime}
\end{align}

The first order SUSY equations \cite{Apruzzi:2013yva} for the fields $\tilde{A}(r)$, $\tilde{\phi}(r)$ and $x(r)$ are
\begin{align}
\partial_r\tilde \phi(r) &= \frac{1}{4}\frac{\e^{-\tilde A(r)}}{\sqrt{1-x(r)^2}}\left( 12x(r)+\left( 2x(r)^2-5 \right) F_0 \e^{\tilde A(r)+\tilde \phi(r)} \right), \label{e:susy_dgln1} \\
\partial_r x(r) &= -\frac{1}{2}\e^{-\tilde A(r)}\sqrt{1-x(r)^2}\left( 4+x(r) F_0 \e^{\tilde A(r)+\tilde \phi(r)} \right),\\
\partial_r \tilde A(r) &= \frac{1}{4}\frac{\e^{-\tilde A(r)}}{\sqrt{1-x(r)^2}}\left( 4x(r) - F_0 \e^{\tilde A(r)+\tilde \phi(r)} \right). \label{e:susy_dgln3}
\end{align}
Using (\ref{e:A_relation})--(\ref{e:rt_dgl_r}) in these equations and in \eqref{e:alpha_r_H}, we obtain three first order equations for $A(\theta)$, $B(\theta)$ and $\phi(\theta)$ and one algebraic equation for $\lambda(\theta)$:
\begin{align}
A^\prime &= \frac{4x - (2x^2-1)F_0 \e^{A+\frac{5}{4}\phi}}{64\sin\theta}, \label{eee}\\
B^\prime &= \frac{36x +(6x^2+1)F_0 \e^{A+\frac{5}{4}\phi} -64\cos \theta}{64\sin\theta}, \label{ttt}\\
\phi^\prime &= \frac{12x +(2x^2-5)F_0\e^{A+\frac{5}{4}\phi}}{16\sin\theta}, \label{uuu} \\
\lambda &= -\frac{1}{F_0}\left(6\e^{-A-\frac{5}{4}\phi} + x F_0\right), \label{jjj}
\end{align}
where $^\prime=\frac{\partial}{\partial \theta}$ and $x$ should now be read as a shorthand for the function 
\begin{equation}
x(r(\theta))=\sqrt{1-16 \e^{2B(\theta)-2A(\theta)} \sin^2\theta}.
\end{equation}
Note that the supersymmetry conditions are chosen such that they are consistent with the choice $F_0 < 0$ if regular boundary conditions (i.e., boundary conditions without brane sources) are imposed at the south pole and with $F_0 > 0$ for the case of regular boundary conditions at the north pole \cite{Apruzzi:2013yva}.

The equations \eqref{eee}--\eqref{jjj} imply that (\ref{e:alphaprime}) follows from (\ref{e:alpha_r_H}) upon differentiation. We thus seem to need four extra conditions 
for a supersymmetric solution, namely the first order equations (\ref{eee})--(\ref{uuu}) as well as the algebraic constraint (\ref{jjj}).
However, taking the derivative of one of the first order equations (\ref{eee})--(\ref{uuu}) and the constraint (\ref{jjj}), the second order
equations (\ref{eoms-theta-dilaton})--(\ref{eoms-theta-inteinsteintrans}) imply the other two first order equations. 
Thus, the supersymmetry equations of \cite{Apruzzi:2013yva} altogether only impose two additional constraints on the functions $A$, $B$ , $\phi$ and $\lambda$ beyond the general field equations of Section \ref{sec:setup}. This is confirmed by evaluating (\ref{eee})--(\ref{uuu}) and (\ref{jjj}) in a series expansion near an anti-D6-brane, as this results in two constraints on the local parameters $a_0,b_0,f_0,\lambda_0,\lambda_1,F_0$, which can be used to eliminate, e.g., $\lambda_0$ and $\lambda_1$,
\begin{equation}
\lambda_0 = - 6\frac{a_0f_0^5}{F_0}, \qquad 
\lambda_1 = \frac{1}{4} \frac{\lambda_0 b_0 F_0 +32 \lambda_0 a_0^3 f_0^5 - 4 a_0b_0f_0^5}{a_0 b_0 f_0^5}. \label{conditions}
\end{equation}

Following our discussion in Section \ref{sec:setup}, our interpretation is that one particular combination  of the constraints ensures that we can integrate all fields to $\theta=\pi$ and that the remaining constraint selects the supersymmetric solution among a one-parameter family of solutions that generically do not satisfy all the supersymmetry equations listed above. This will be confirmed by our numerical considerations in the next section.

\subsection{The D8-brane potential in the supersymmetric case}

Using \eqref{conditions} in \eqref{pol-coefficients}, we find that the coefficients $c_{2}$, $c_{3}$, $c_{4}$ of the D8-brane potential simplify to
\begin{equation}
c_{2} = -\frac{4}{a_0^5 b_0 f_0^3}, \qquad c_{3} = -\frac{2}{a_0^6b_0^{3/2}f_0^5}, \qquad c_{4} = \frac{1}{2a_0^7b_0^2f_0^7}. \label{susy-coefficients}
\end{equation}
We immediately see that the quadratic coefficient $c_2$ is manifestly negative so that the anti-D6-branes do polarize in the presence of the AdS curvature. This is one of our main results. In order to further analyze the potential \eqref{d8potential}, we introduce the shorthand
\begin{equation}
\bar{\theta}:= \frac{1}{n a_0b_0^{1/2}f_0^2} {\theta}
\end{equation}
in terms of which the potential takes the simple form
\begin{equation}
V(\bar\theta) \propto n^3 \bar{\theta}^2 \left(-2 + 2\bar{\theta} +\bar{\theta}^2\right).
\end{equation}
The extrema are at $\bar{\theta}=0$ and at $\bar{\theta}=\frac{1}{2}$. The latter extremum is the minimum and corresponds to
\begin{equation}
\theta=\theta_{\star}=\frac{na_0b_0^{\frac{1}{2}}f_0^2}{2}.
\end{equation}
As explained in Section \ref{regime}, it is always possible to adjust the parameters such that the minimum is at small $\theta$ and all other approximations used to derive the potential are justified. Using \eqref{susy-coefficients} in \eqref{mass}, we furthermore find $m^2 = -8$ for the squared mass of $\theta$. As expected for a supersymmetric solution, this is above the BF bound such that brane polarization happens non-perturbatively via tunnelling to the lower-lying minimum at finite $\theta$.

\begin{table}[t!]
  \centering
  \begin{tabular}{cc|rrrrr}
    \hline
    \multicolumn{2}{c}{Initial Values}&\multicolumn{5}{c}{Values at the South Pole}\\
    $F_0$&$\tilde \alpha_0$&$a_\mathrm{s0}$&$b_\mathrm{s0}$&$\alpha_\mathrm{s0}$&$f_\mathrm{s0}$&$\alpha_\mathrm{s2}$\\
    \hline
    4, Susy &  2.5& 	  0.8887& 	 10.2009& 6.0000& 	  1.4038& 0.0741\\
    40, Susy&       25&   0.4821&   1.6871&   60.000&   2.2413&   2.8210\\
    10& 6.93183447&     0.5950&     1.4746&    20.2938&     1.4050&      6.5964\\
    15& 10.51328&	0.5126&   0.6110&  31.2296&   1.3966&  37.1415\\
    40& 28.247186&    0.3668&    0.0929&   83.5936&    1.3583& $1.14\cdot 10^3$\\
    \hline
  \end{tabular}
  \caption[Numerically Calculated Coefficients at the End of the Solution for Given Initial Values]{Initial values and coefficients at the south pole for solutions with integration range $[0,\pi]$ and no source at the north pole, where $\tilde \alpha_0=\tilde \lambda_0 \tilde a_0^{-7} \tilde f_0^{-3}$ denotes the initial value of $\alpha(\theta)$ at $\theta=0$ and the subscript s refers to coefficients at the south pole. All solutions except for the $F_0=40$ Susy solution have initial values $\tilde a_0 =\tilde  b_0 =\tilde  f_0 = 1$. For the $F_0=40$ Susy solution, $\tilde a_0 = 10^{-\frac{1}{4}}$, $\tilde b_0=1$ and $\tilde f_0=10^{\frac{1}{4}}$ was chosen. The given number of digits is necessary to get a maximal range of at least $\theta_\textrm{int} = 3.1415$.}
\label{t:para_pi}
\vspace{1em}
\end{table}

\section{Non-supersymmetric solutions?}
\label{sec:non-susy}

In Section \ref{sec:setup}, we discussed how many independent parameters are needed to specify a general solution to the second order field equations 
(\ref{eoms-theta-dilaton})--(\ref{eoms-theta-inteinsteintrans}).
The original six-dimensional parameter space spanned by $a_0,b_0,f_0,\lambda_0,\lambda_1,F_0$ is reduced to a 
three-dimensional hypersurface when the three physical parameters
$F_0,Q_1,Q_2$ are fixed. Furthermore, $b_0$ parameterizes a residual coordinate freedom,
leaving a two-dimensional hypersurface of gauge orbits in the parameter space.
The supersymmetric solutions discussed in the previous section come with two additional
constraints on the boundary values and hence correspond to a point in this two-dimensional hypersurface of gauge orbits.

\begin{figure}[t!]
\centering
  \includegraphics[scale=1.00]{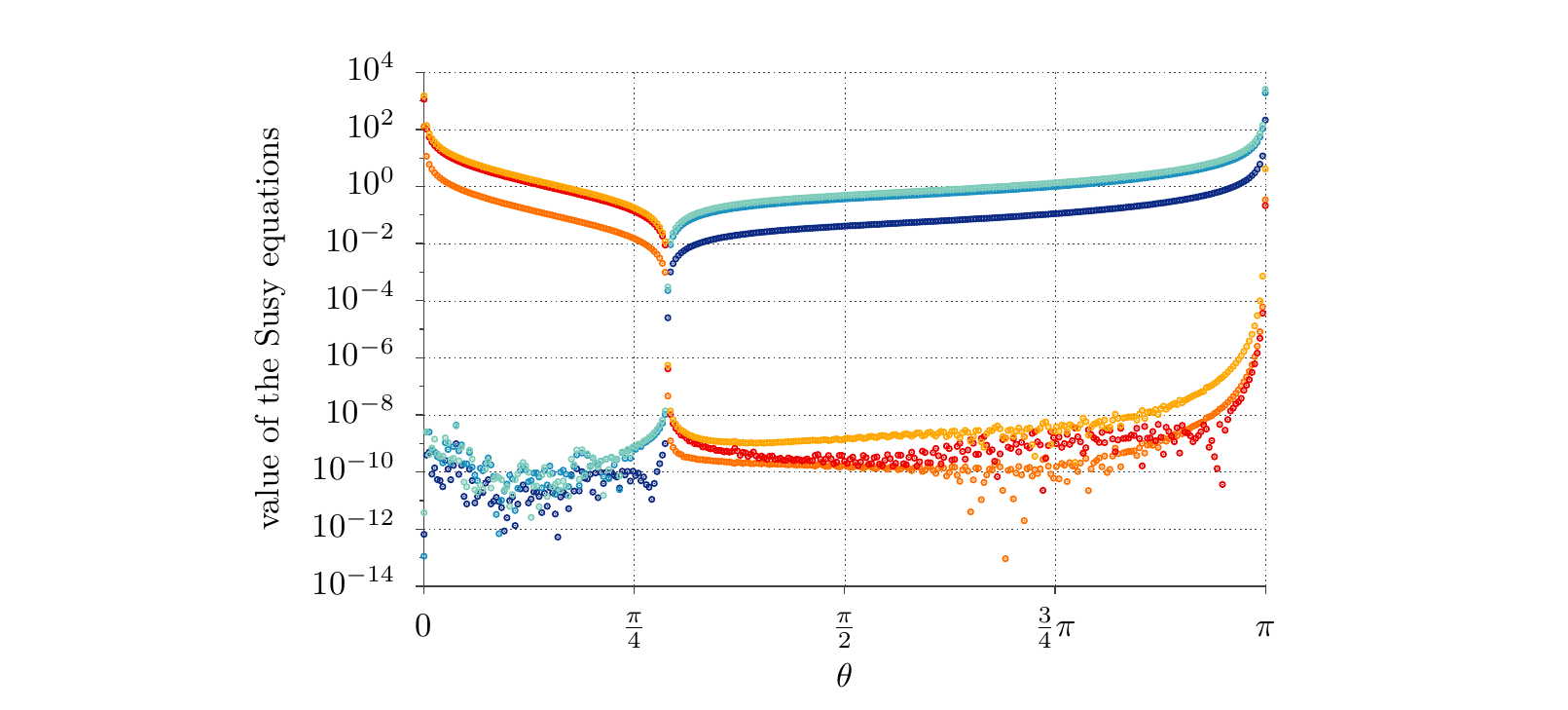}
  \caption[Supersymmetric Equations Evaluated for a Supersymmetric Solution]{The three supersymmetric differential equations \eqref{eee}--\eqref{uuu} evaluated for the general solution with $\tilde a_0 = 1, \tilde b_0 = 1, \tilde f_0 = 1, \tilde \lambda_0 = -2.5, F_0 = 4$ for positive $x(\theta)$ (light, medium and dark blue) and negative $x(\theta)$ (yellow, orange and red), where the value $0$ means that an equation is solved. One can see that the equations are fulfilled when $x(\theta)$ starts positive at $\theta=0$ and then switches its sign between the two poles. This is consistent with the conventions of \cite{Apruzzi:2013yva}, where $x=1$ at the north pole and $x=-1$ at the south pole.}
  \label{f:susydiff}
\vspace{1em}
\end{figure}

\begin{figure}[t!]
  \centering
  \includegraphics[scale=1.00]{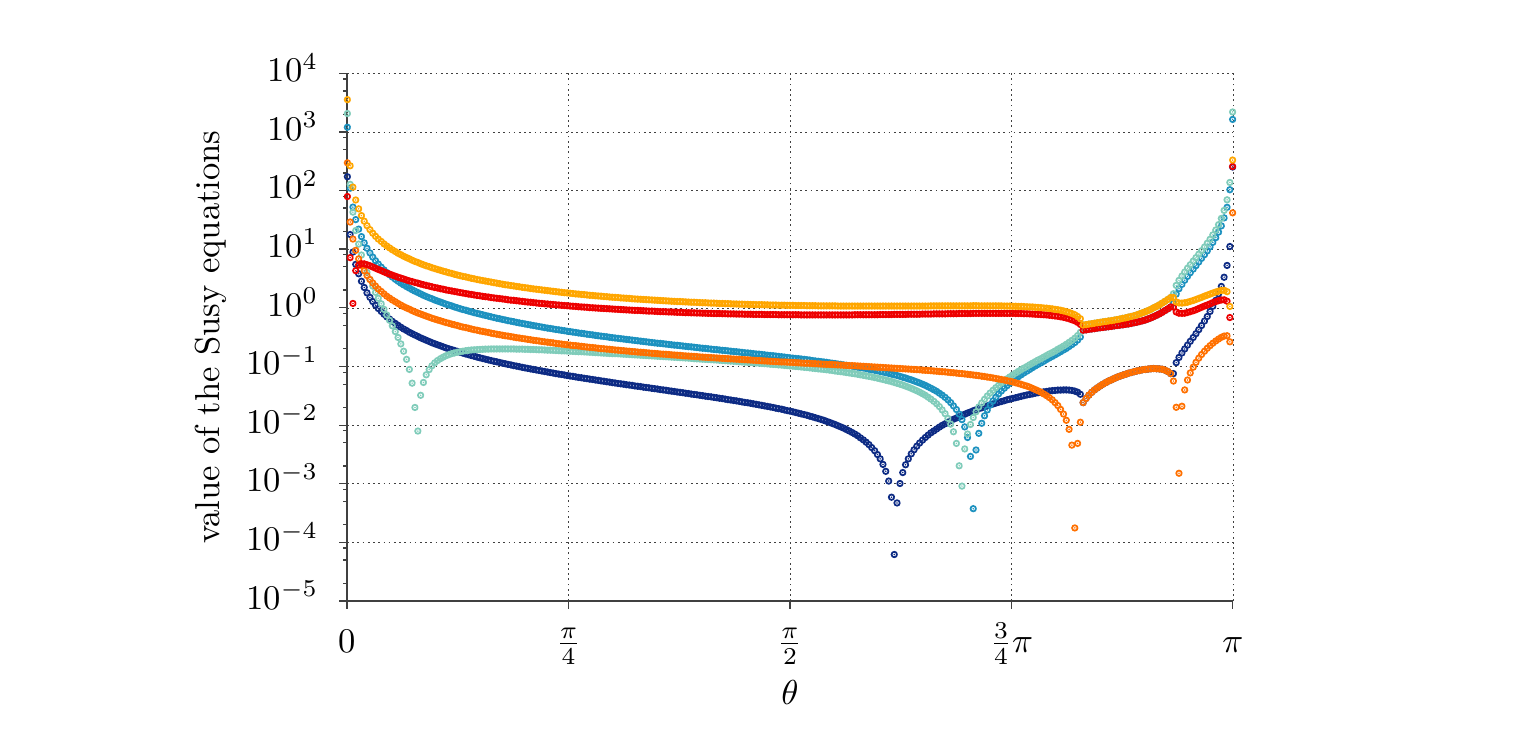}
  \caption[Supersymmetric Equations Evaluated for a Non-supersymmetric Solution]{The three supersymmetric differential equations evaluated as in \autoref{f:susydiff} for the solution with $F_0 = 15$ and $\tilde \alpha_0$ from \autoref{t:para_pi} to get maximal integration range. As the equations have deviations from zero ranging between $10^{-3}$ and $10$, they are not well fulfilled in comparison to the supersymmetric solution in \autoref{f:susydiff}. In addition, all deviations are large for $\theta \approx 0$ and there is no clear point at which the solution of $x$ switches the sign. This supports the argument that this solution is not supersymmetric although it can be integrated over the whole range.}
  \label{f:nosusydgldiff}
\vspace{1em}
\end{figure}

In this section we would like to explore whether this supersymmetric point is the only possible solution to the 
full second order field equations, or whether the other points on the two-dimensional hypersurface of gauge orbits
also contain meaningful solutions. To this end, we considered numerical solutions of the second order field equations in the case when there are no anti-D6-branes
at the north pole $\theta=0$, i.e., the boundary condition at the north pole is given by \eqref{eq:d6-ansatz-bc2} and all possible anti-D6-branes are concentrated at the south pole at $\theta=\pi$. This  is done in order to have a smooth starting point for the numerics (see also \cite{Schmidt:2013} for details on the numerical treatment).
Fixing furthermore $\tilde a_0=\tilde f_0=1$ (which can always be achieved by the rescaling symmetries (\ref{scale})) and choosing a fixed value $\tilde b_0=1$, one finds that the numerically computed solutions become singular well before $\theta=\pi$ is reached unless one restricts oneself to a certain one-dimensional subspace in the $(\tilde \lambda_0,F_0)$-plane, on which the field equations can be integrated up
to $\theta\approx \pi$ to very good approximation. If our reasoning is correct, only one point on this line should correspond to a supersymmetric solution as described in Section \ref{sec:susy}.
This is confirmed by evaluating the first order susy equations (\ref{eee})--(\ref{uuu}) on those numerical solutions that integrate up to $\pi$. Writing them in the form $0=\ldots$, one 
obtains values for the right-hand sides between $10^{-3}$ and $10$, which should be compared with values between $10^{-10}$ and $10^{-8}$ that one finds for supersymmetric parameter
sets. This is illustrated in Table \ref{t:para_pi}, which contains two supersymmetric and three non-supersymmetric pairs of initial values $(\tilde \lambda_0,F_0)$ for which the solutions integrate up to $\theta=\pi$ (for the supersymmetric solutions, this is always the case). Figures \ref{f:susydiff} and \ref{f:nosusydgldiff} show how well and how badly the supersymmetry equations (\ref{eee})--(\ref{uuu}) are fulfilled for a supersymmetric and a non-supersymmetric solution, respectively \cite{Schmidt:2013}.

\begin{figure}[t!]
  \centering
  \input{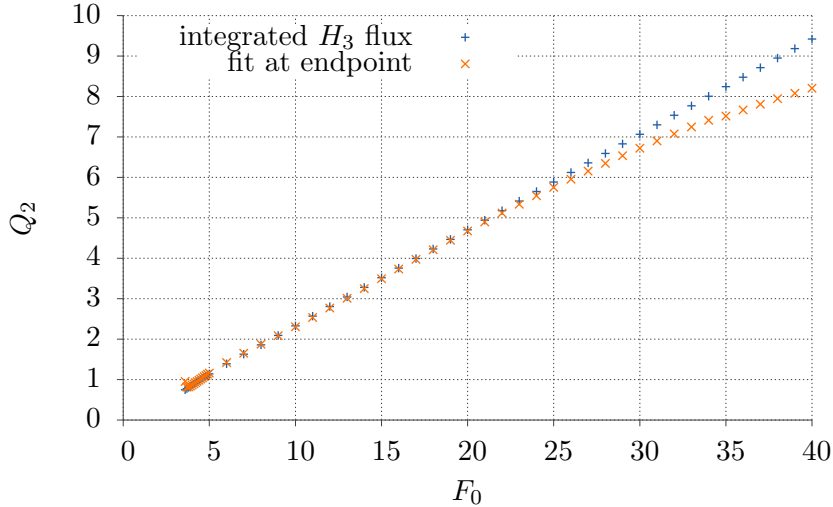}
  \caption[The integrated Bianchi identity]{The figure shows Matlab results for the integrated $F_2$ Bianchi identity for different values of $F_0$ and $Q_2$. In order that the D6 tadpole is cancelled in a numerical solution, the charge obtained from the integrated $H_3$ flux (blue dots) must equal the charge from the delta function source term, which is read off from a fit of the solution at the south pole (orange dots). Except for the dot at $F_0=4$, all solutions are non-supersymmetric, and all integrate to $\theta\approx\pi$. The main numerical uncertainty comes from the tuning necessary to get the integration range until $\theta=\pi$, which is much easier for supersymmetric solutions. Nevertheless, the non-supersymmetric solutions satisfy the integrated Bianchi identity very well. For the larger values of $F_0$, numerical problems due to small numbers become relevant.}
  \label{f:Bianchiidentity}
\vspace{1em}
\end{figure}

If these non-supersymmetric solutions are to be physically meaningful, they should give rise to physically sensible boundary conditions at the south pole, i.e., they should asymptote to the anti-D6-brane boundary conditions with some brane charge $Q_2$ and also satisfy the integrated Bianchi identity for $F_2$ that corresponds to this charge. In order to be able to read off a charge $Q_2$ from the asymptotics at the south pole, however, one first has to scan through the parameter space to find a suitable pair $(\tilde \lambda_0,F_0)$ that integrates to $\theta=\pi$ with sufficient numerical accuracy. This is the primary obstacle for identifying good non-supersymmetric solutions, but once this is achieved, the charge $Q_2$ can be read off and compared with the integrated Bianchi identity. In Figure \ref{f:Bianchiidentity}, we show the results of this comparison for a set of non-supersymmetric solutions. As is shown there, the Bianchi identity is fulfilled to good accuracy, especially for parameter regimes with reduced numerical problems.

\begin{figure}[t!]
  \centering
  \input{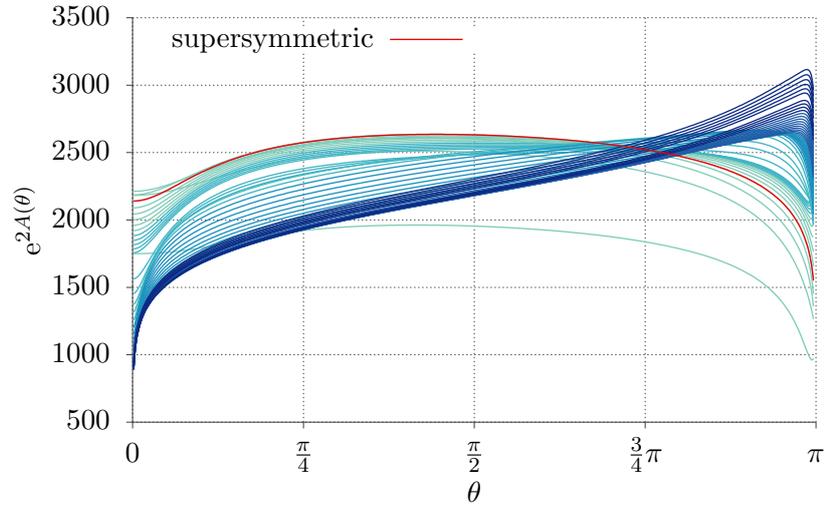}
  \caption[Warp factor]{The function $\e^{2A(\theta)}$ for a set of solutions with fixed $(F_0,Q_2)$.}
  \label{f:warp}
\vspace{1em}
\end{figure}

\begin{figure}[t!]
  \centering
  \input{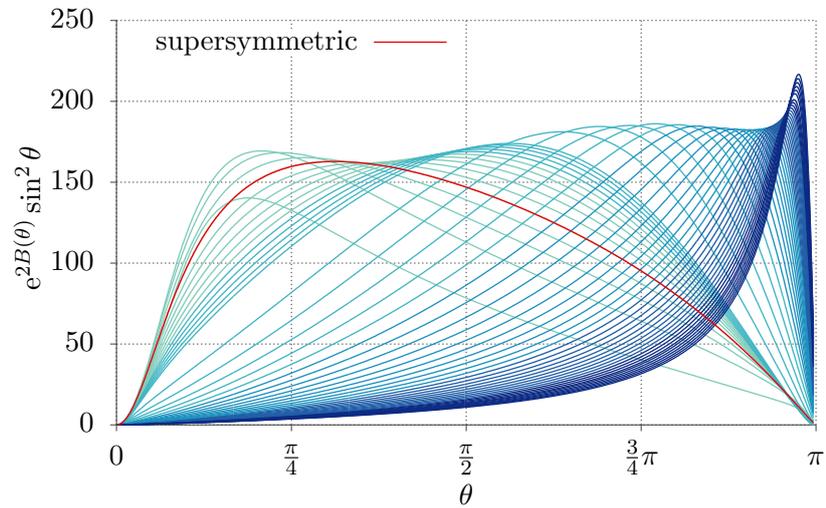}
  \caption[2-sphere radius]{The function $\e^{2B(\theta)}\sin^2\theta$ measuring the squared 2-sphere radius for a set of solutions with fixed 
  $(F_0,Q_2)$.}
  \label{f:twosphere}
\vspace{1em}
\end{figure}

We also searched for a set of non-supersymmetric solutions and a supersymmetric solution that all correspond to the same physical parameters $(F_0,Q_2)$, by fitting the 
initial values accordingly. The profiles of the various functions $A$, $B$, $\phi$, $\alpha$ for the choice $F_0=4$, $Q_1=0$, $Q_2=400$ are displayed in Figures \ref{f:warp} to \ref{f:alpha}.

\begin{figure}[t!]
  \centering
  \input{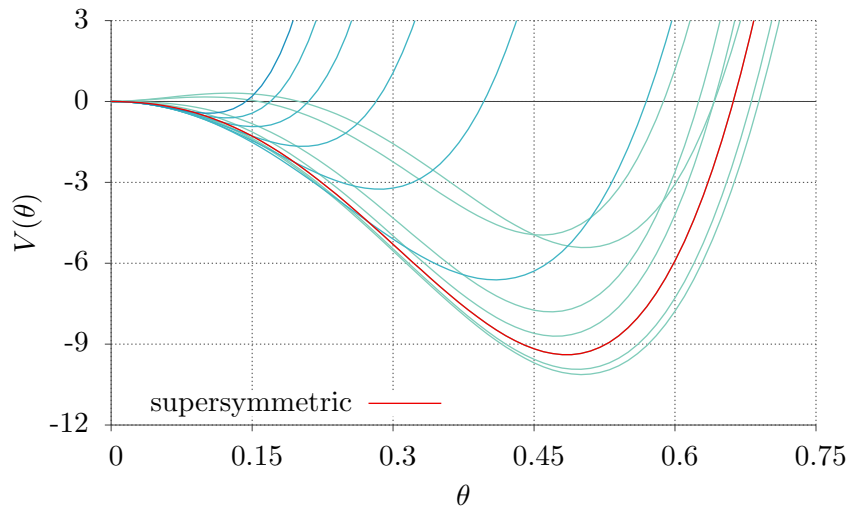}
  \caption[potential]{The D8-brane potential as a function of $\theta$ for a set of solutions with fixed $(F_0,Q_2)$. All plotted solutions have a minimum at finite $\theta$ which is lower than the extremum at the origin and therefore allow a polarization into a D8-brane.}
  \label{f:potential}
\vspace{1em}
\end{figure}

\begin{figure}[t!]
  \centering
  \input{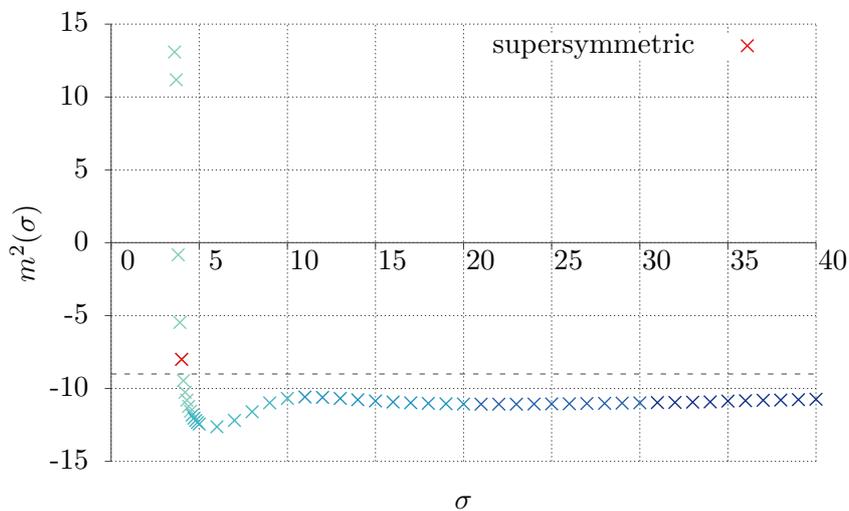}
  \caption[msquare]{The squared mass of the worldvolume scalar $\theta$ for several non-supersymmetric and one supersymmetric solution, parametrized by $\sigma$ (in arbitrary units). At the supersymmetric point and in its vicinity, $m^2$ is negative but above the BF bound $m^2=-9$ (broken line). Further to the right, the solutions become tachyonic, whereas $m^2$ becomes positive left from the supersymmetric solution.}
  \label{f:msquare}
\vspace{1em}
\end{figure}

Putting all this together, we find numerical evidence for a one-parameter family of solutions that are non-supersymmetric except for a single parameter value, which corresponds to the supersymmetric solution 
found in \cite{Apruzzi:2013yva}.
More precisely, the non-supersymmetric solutions satisfy the second order field equations to a comparable accuracy as the supersymmetric solution, but they clearly
violate the first order supersymmetry equations \eqref{eee}--\eqref{uuu}. We performed several numerical tests that could have exposed these non-supersymmetric solutions as numerical artifacts, but
none of them provided any evidence in this direction.
Assuming these solutions to exist, we then studied the corresponding polarization potential and found several regimes with a different qualitative behavior. For non-supersymmetric solutions that lie in the vicinity of the supersymmetric one in moduli space, the D8-brane potential has the same qualitative features of a maximum at the origin and a minimum at finite $\theta_{\star}$. This is shown in Figure \ref{f:potential} for the choice $F_0 =4$, $Q_1=0$, $Q_2=400$, $n=20$. We also computed the value of $m^2$ at the origin and found that the BF bound is satisfied in the region close to the supersymmetric solution (cf. Figure  \ref{f:msquare}). Further away from the supersymmetric point, however, we also found two other regimes, which are qualitatively different from the first one. Moving away from the supersymmetric point in one direction of the parameter $\sigma$ (cf. Section \ref{sec:parameters}), $m^2$ becomes more and more negative and eventually violates the BF bound such that the worldvolume scalars become tachyonic and brane polarization can happen already perturbatively. On the other hand, if one deviates from the supersymmetric solution in the other direction, $m^2$ approaches zero and eventually becomes positive such that the maximum at the origin of the polarization potential becomes a minimum. As shown in Figure \ref{f:potential}, this minimum is then separated from a second, lower-lying minimum at finite $\theta$ by a maximum such that brane polarization again happens non-perturbatively in these solutions.

We should stress again that, since the non-supersymmetric solutions have been obtained numerically, we cannot fully exclude that their different $\theta$ masses are a numerical artifact. While our numerical data does not suggest that this is the case, it would nevertheless be important in future work to obtain an analytical understanding of the parameter $\sigma$ that scans the different solutions.

\section{Conclusions}
\label{sec:conclusions}

In this paper, we have linked the supersymmetric first order solution of massive type IIA compactifications on AdS$_7\times S^3$ with anti-D6-branes and oppositely charged $H_3$ flux of \cite{Apruzzi:2013yva} to the general framework of second order solutions of this setup discussed in \cite{Blaback:2011nz, Blaback:2011pn}. Using the extra constraints imposed by supersymmetry, we were able to compute the potential of a probe D8-brane with dissolved anti-D6-brane charge in a background with fully backreacted anti-D6-branes. This complements a similar computation in \cite{Bena:2012tx}, where the same potential was determined in a non-compact version of the setup. As we showed in this paper, the D8-brane potential in the compact case has a universal behavior with a local maximum at the origin and a local minimum at a finite angular distance away from it. This suggests that the anti-D6-branes polarize, as opposed to their counterparts in the non-compact setup. The difference between the two cases can be traced back 
to the curvature of the external spacetime, which is required to be negative in the compact case due to the integrated Einstein equations, by which it is tied to the energy of the field strengths in the compact dimensions. In the non-compact case, no such constraint on the external curvature exists such that one has the freedom to take the external spacetime to be Minkowski as in \cite{Bena:2012tx}. The results obtained in this paper thus show that compactification effects can resolve flux singularities.

This interpretation is consistent with the recent independent results in version 2 of \cite{Apruzzi:2013yva}, where a configuration with D8-branes wrapping a finite $S^2$ was found using the supergravity equations. Our results suggest that the solution found in \cite{Apruzzi:2013yva} is in fact the stable end configuration that replaces the singular configuration found in \cite{Blaback:2011pn}.

We also explored the solution space beyond the supersymmetric subspace identified in \cite{Apruzzi:2013yva} and found numerical evidence for a one-parameter family of apparently non-supersymmetric solutions. While our understanding of the physical parameter or modulus that scans these solutions is incomplete, we computed numerically the D8-brane potential also for these cases and found again that the anti-D6-branes tend to polarize.

Our results raise several interesting questions for future research, e.g., regarding the CFT-dual of the final configuration with D8-branes \cite{Gaiotto:2014lca} or the role of curvature and compactness for brane polarization. It would also be interesting to see whether there is any lesson to be learned for the anti-D3-brane singularity of KKLT-like setups or to explore possible connections to the recent results of \cite{Bena:2014bxa} on anti-M2-branes.

\section*{Acknowledgements}

The authors would like to thank Fabio Apruzzi, Iosif Bena, Stanislav Kuperstein, Stefano Massai and Thomas Van Riet for useful discussions. DJ also thanks the Institute for Theoretical Physics at Leibniz Universit{\"{a}}t Hannover for hospitality during a visit. This work was supported in part by the German Research Foundation (DFG) within the Cluster of Excellence ``QUEST''.

\clearpage

\begin{figure}[t!]
  \centering
  \input{tcoupling.tex}
  \caption[Dilaton]{The function $\e^{\phi(\theta)}$ for a set of solutions with fixed 
  $(F_0,Q_2)$.}
  \label{f:dilaton}
\vspace{1em}
\end{figure}

\begin{figure}[ht!]
\vspace{1em}
  \centering
  \input{talpha.tex}
  \caption[alpha]{The function $\alpha(\theta)$ for a set of solutions with fixed 
  $(F_0,Q_2)$.}
  \label{f:alpha}
\vspace{1em}
\end{figure}

\clearpage

\bibliographystyle{utphys}
\bibliography{groups}

\end{document}